# Experimental scaling of the scrape-off layer particle flux width by outboard divertor Langmuir probes with favorable $B_t$ configuration on EAST

X. Liu[1], L. Y. Meng[1], J. C. Xu[2], L. Wang[1, 3, *], J. Li[1, 3], and the EAST Team

[1]*Institute of Plasma Physics, Chinese Academy of Sciences, Hefei 230031, People's Republic of China*

[2]*School of Mechanical Engineering, Anhui University of Science & Technology, Huainan 232001, People's Republic of China*

[3]*Institute of Energy, Hefei Comprehensive National Science Centre, Hefei 230031, People's Republic of China*

[*]E-mail: lwang@ipp.ac.cn

**Abstract**

The scrape-off layer (SOL) power width ($\lambda_q$) is important for predicting the heat load on divertor targets for future magnetically confined devices. Currently, the underlying physics for $\lambda_q$ scaling is not fully understood. This paper extends the previous inboard SOL particle flux width ($\lambda_{js}$, $\lambda_q \approx \lambda_{js}$ is assumed) scaling [Liu *et al* 2019 *Plasma Phys. Control. Fusion* 61 045001] to the outboard side in EAST, which can provide more experimental evidence for $\lambda_q$ study. A systematic method has been developed to correct the less reliable upper outer (UO) divertor Langmuir probe (Div-LP) measurements with their more reliable neighboring measurements to reduce the measurement uncertainty of $\lambda_{js}$. For the discharges with the favorable $B_t$ and upper single null configurations in the 2019 experiment campaign, about 260 discharges have been selected by certain criteria to ensure good $\lambda_{js}$ measurements. Three H-mode, L-mode, and Ohmic databases have been constructed and are used for $\lambda_{js}$ scalings. It is found that the outboard $\lambda_{js}$ for the H-mode and L-mode plasmas scales as, $\lambda_{js,UO} = 1.52(W_{MHD}/\bar{n}_e)^{-0.61}P_{tot}^{0.19}$, where $W_{MHD}$ is the stored energy, $\bar{n}_e$ is the line-averaged density, and $P_{tot}$ is the total input power. This scaling is similar to the inboard $\lambda_{js}$ scaling except for the scaling amplitude that is probably due the triangularity. The repeatable scaling dependence on $W_{MHD}/\bar{n}_e$ confirms the reliability of this dependence even though the regression quality is relatively poor. It is also discussed that the solely scaling of $\lambda_q/\lambda_{js}$ on $B_p$ is not enough to include all the physics of SOL heat transports.

## 1. Introduction

It is well known that the divertor heat load is a crucial problem for future magnetically confined fusion devices. The mitigation of the excessive heat load on divertor targets needs a better understanding of the particle and heat transports in the edge and scrape-off layer (SOL). The SOL power width ($\lambda_q$) is an important parameter to characterize the SOL heat transports. $\lambda_q$ is determined by the competition between the radial and parallel heat transports in the SOL. Many experimental, theoretical, and numerical studies [1-15] on $\lambda_q$ have been carried out to understand the underlying physics that determines $\lambda_q$. However, there are still many open questions, e.g., whether the radial heat transport in the SOL is dominated by the magnetic drift, and if $\lambda_q$ has a strong scaling dependence on the machine size? So, more experimental results are required to validate the simulation results and understand the dominant physics for $\lambda_q$ scaling.

In the previous work, the SOL particle flux width ($\lambda_{js}$) measured by the upper-inner (UI) divertor Langmuir probes (Div-LPs) for H-mode plasmas was found to scale as,

$$\lambda_{js,UI,H} = 0.85(W_{MHD}/\bar{n}_e)^{-0.50} P_{tot}^{-0.06}, \qquad (1)$$

where $W_{MHD}$ is the stored energy, $\bar{n}_e$ is the line-averaged density, and $P_{tot}$ is the total input power [7]. Normally, the outboard measurements by Div-LP, thermocouple, and IR camera are used to scale $\lambda_q/\lambda_{js}$. However, EAST's plasma operations regularly utilize upper single null (USN) geometry, forward (unfavorable) toroidal magnetic field ($B_t$) direction, and low hybrid wave (LHW) heating scheme. The upper outer (UO) strike point normally splits [16], making it difficult to evaluate the outboard $\lambda_q/\lambda_{js}$. Although the inboard Div-LP data were used as an alternative to scale $\lambda_{js}$, the outboard $\lambda_{js}$ scaling might be different due to the in/out asymmetry. Furthermore, the scaling dependence of inboard $\lambda_{js}$ on $W_{MHD}/\bar{n}_e$ needs to be confirmed, since the regression quality is relatively poor (see figure 5(b) in reference [7]). In EAST's 2019 experiment campaign, there were plenty of discharges configured with reversed (favorable) $B_t$ direction. Then, the UO Div-LP data can be used to scale the outboard $\lambda_{js}$. The rest of this paper is organized as follows: section 2 introduces the basic experimental conditions; section 3 describes the correction of the UO Div-LP measurements; section 4 introduces the scaling of outboard $\lambda_{js}$ and its comparison with previous studies; section 5 summarizes the whole paper.

## 2. Experiment conditions

The basic machine parameters of EAST [18], the divertor shape and materials (ITER-like tungsten/copper divertors [19] for the upper divertors), and the employed main diagnostics (Div-LPs [20]) are described in section 2.1 of reference [7]. The UO Div-LPs have two toroidally symmetric arrays, locating at the D and O ports (toroidal displacement is 112.5°). The discharges in the 2019 experiment campaign that were configured with USN geometry and reversed (favorable) $B_t$ direction are chosen to avoid strike point splitting at the UO divertors. The shot number ranges from #86377 to #88018. Note that from #86790 to #87475, most of the discharges were helium (or mixed with deuterium) plasmas. To ensure the consistency of the correction of the Div-LP measurements in the next section, these discharges are included. But the analysis of the Div-LP data for helium plasmas is still in progress. The discharges are classified into three types: the Ohmic, L-mode, and H-mode discharges. For the Type-I ELMy H-mode discharges, only the inter-ELM Div-LP data are retained. The heating scheme typically utilized radio-frequency waves (LHW and/or electron cyclotron resonance heating). The plasma was fueled with supersonic molecular beam injection and the wall was coated with lithium. The discharges with divertor impurity seeding or resonant magnetic perturbation are omitted.

## 3. Correction of the divertor Langmuir probe measurements

The evaluation of $\lambda_{js}$ by EAST's upper Div-LPs normally has large uncertainty, due to the relatively sparse probe distribution in the poloidal direction and the large measurement uncertainty of the ion saturated current density ($j_s$) when the probe tip is eroded. To overcome these difficulties, several methods have been utilized. In the previous work, the combination of the UI Div-LP measurements and the selection of the fitted $j_s$ profile are employed to increase the measurement reliability of $\lambda_{js}$ for the sparse Div-LPs [7]. An experimental scaling has been proposed to calibrate the collecting area of the eroded Div-LP tip near the strike point to reduce the measurement uncertainty of $j_s$ [17]. This scaling is useful for Div-LP calibration only in the early stage of one experiment campaign, where the erosion of Div-LP tip is not severe (see figure 5 in reference [17]). For the chosen discharges in section 2, there were about 4000 discharges ahead in this campaign. The UO Div-LP tips near the strike point had already been seriously eroded. So, the method in reference [17] is not applicable. In this section, a new method is proposed to correct the $j_s$ measurements by Div-LPs with the following assumptions: i) the profile of the parallel particle flux ($\Gamma = j_s/e$, where $e$ is the elementary charge) at the divertor mapped to the outboard midplane (OMP) can be well fitted by the Eich function (see equation (2) in reference [7], which is modified from equation (1) in reference [2]) for Ohmic discharges; ii) the profile of $\Gamma$ is toroidally symmetric in Ohmic discharges; iii) there are no 3D strike point splitting at the UO divertors for the chosen discharges in section 2. The conditions to use this method rely on that the profile width of $\Gamma$ is large enough to be well captured by Div-LPs and most of Div-LPs are in good condition, especially the neighboring channels of the channel to be corrected.

The basic idea of this correction method is to evaluate the reliability of a Div-LP channel by comparison of the measurements with its neighboring channels and correct the measurement of a relatively unreliable channel (not faulty, reliable in a short period) with the reliable ones (reliable in a long period). As mentioned in section 2, there are two toroidally symmetric Div-LP arrays at the UO divertors i.e., the UO-D and UO-O Div-LPs. This is beneficial to the correction process, as it increases the number of Div-LP channels to be referenced. Figure 1 shows the measured $j_s$ profiles (the time average range is 4 s ± 50 ms) by the UO-D and UO-O Div-LPs mapped to the OMP (by EFIT equilibrium) for the first Ohmic discharge (#86568) regarding the chosen discharges. The channel number (Ch.) is shown next to the corresponding measurement. For a UO-D Div-LP with channel number of $N$ ($N$ = 1, 2…, 13), we name the UO-O Div-LP with channel number of $N$+13 as its twin channel and vice versa. The twin channels have the same value of $R$-$R_{LCFS}$ ($R_{LCFS}$ is the major radius of the last closed flux surface (LCFS) on the OMP obtained from EFIT) and they are important reference channels for each other. The measurements from channels 1, 2, and 11 are not trustable (faulty in this discharge) and are removed in the evaluation of $\lambda_{js}$. They are shown just for clarification. The red solid line represents the Eich fit of the UO-O Div-LP measurements. With the assumptions mentioned above, the measurements from the UO-O Div-LPs (like channels 14 and 18) are slightly adjusted to match the Eich fit. A new fit is then performed for the UO-O Div-LP measurements and the UO-D Div-LP measurements that are biased from the Eich fit (not for faulty channels) are also adjusted. The reason to make the UO-D Div-LP measurements to align with the UO-O Div-LP measurements is that the UO-D Div-LPs have faulty channels occasionally (like channels 1, 2, and 11) and their measurements are less reliable (will be demonstrated later).

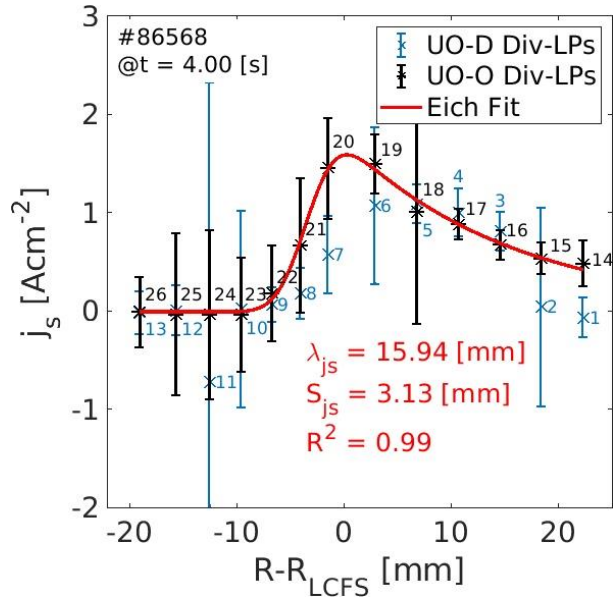

Figure 1 The measured $j_s$ profiles by the UO-D and UO-O Div-LPs. The profiles are mapped to the outboard midplane by EFIT equilibrium. $R_{LCFS}$ is the major radius of the last closed flux surface. The red solid line represents the Eich fit of the UO-O Div-LP measurements. The channel number is shown next to the corresponding measurement.

The reliability of one channel is evaluated by calculating its Pearson correlation coefficients (Corr) with its neighboring channels, its twin channel, and the neighboring channels of its twin channel. To evaluate the overall reliability, the Corr is calculated by concatenating $j_s$ measurements for all valid discharges of the chosen discharges in section 2, but the sampling rate is reduced to 1 Hz (to reduce the amount of data). The results are listed in table 1. The second column is the Corr between twin channels. The third and sixth columns are the Corr between one channel and the neighboring channel of its twin channel. The fourth and fifth columns are the Corr between neighboring channels. The seventh and eighth columns are the mean Corr for the UO-D and UO-O Div-LPs, respectively. The mean Corr for one channel is the average Corr calculated with its neighboring channels, its twin channel, and the neighboring channels of its twin channel. Since channels 5 and 18 are close to the strike point for most of the discharges in the 2019 campaign, the probe heads had been seriously eroded. So, their mean Corr are relatively small and the mean Corr for their neighboring channels are also influenced. The first half of the UO Div-LPs regarding the channel number normally lay inside the SOL, which are important for the evaluation of $\lambda_{js}$. Thus, the UO-O Div-LPs are relatively reliable than the UO-D Div-LPs according to table 1.

Table 1. The overall reliability of the UO Div-LPs evaluated by Pearson correlation coefficient (Corr).

| $N$ | Corr(Ch. $N$, Ch. $N$+13) | Corr(Ch. $N$, Ch. $N$+14) | Corr(Ch. $N$, Ch. $N$+1) | Corr(Ch. $N$+13, Ch. $N$+14) | Corr(Ch. $N$+13, Ch. $N$+1) | Mean Corr Ch. $N$ | Mean Corr Ch. $N$+13 |
|---|---|---|---|---|---|---|---|
| 1 | 0.3828 | 0.3810 | 0.4017 | 0.9539 | 0.5772 | 0.3885 | 0.6380 |
| 2 | 0.6057 | 0.5604 | 0.8320 | 0.9498 | 0.8604 | 0.5954 | 0.7502 |
| 3 | 0.8475 | 0.5672 | 0.8801 | 0.6804 | 0.8905 | 0.7974 | 0.7857 |
| 4 | 0.6849 | 0.6091 | 0.3657 | 0.3147 | 0.4601 | 0.6861 | 0.5415 |
| 5 | 0.1072 | 0.3640 | 0.3981 | 0.1185 | 0.0790 | 0.3390 | 0.2457 |
| 6 | 0.8033 | 0.1632 | 0.2828 | 0.0951 | 0.2468 | 0.3453 | 0.3255 |
| 7 | 0.5486 | 0.7242 | 0.6824 | 0.2344 | 0.2307 | 0.4969 | 0.2544 |
| 8 | 0.7471 | 0.7635 | 0.6170 | 0.7467 | 0.4708 | 0.6081 | 0.5846 |
| 9 | 0.6668 | 0.7693 | 0.6825 | 0.7629 | 0.7031 | 0.6413 | 0.7286 |
| 10 | 0.8225 | 0.2827 | 0.5539 | 0.2729 | 0.4797 | 0.6089 | 0.6215 |
| 11 | 0.3122 | 0.5981 | 0.5354 | 0.4075 | 0.3115 | 0.4958 | 0.3174 |
| 12 | 0.8714 | 0.6392 | 0.4977 | 0.6246 | 0.4638 | 0.5710 | 0.5931 |
| 13 | 0.4375 | | | | | 0.4663 | 0.5671 |

Note: The mean Corr for one channel is the mean value of Pearson correlation coefficients calculated with its neighboring channels, its twin channel, and the neighboring channels of its twin channel.

Known the overall reliability, we focus on correcting channels that have small mean Corr (not faulty channels). It should be emphasized that the channel with a small overall mean Corr should be corrected only if its measurements are systematically biased from the measurements by its reliable twin and/or neighboring channels in a certain period. This is explained by an example in figure 2, which shows the concatenated $j_s$ measured by a pair of twin channels. The data are normalized to their first elements. Starting from #87782 (indicated by the dashed red line), the measurements by channel 17 are much lower than those by channel 4 in figure 2(a). If the mismatched measurements by channel 17 are multiplied by a constant (mimicking the correction of its calibration coefficient), the consistency recovers (see the red shaded area in figure 2(b)) and the Corr between these twin channels increases from 0.68 to 0.95. To correct channel 17 (channel 4 is overall reliable according to table 1), the nearest Ohmic discharge (#87842) after #87782 is selected, a similar figure like figure 1 is plotted for this Ohmic discharge, the measurement by channel 17 is aligned with the Eich fit (assuming all other less reliable UO-O channels have been corrected), and the calibration coefficient of channel 17 is updated and is used starting from #87782. Here we name #87782 as the correction point for channel 17. For the overall less reliable channels (like channels 5 and 18), the correction procedure is cumbersome (there are several correction points).

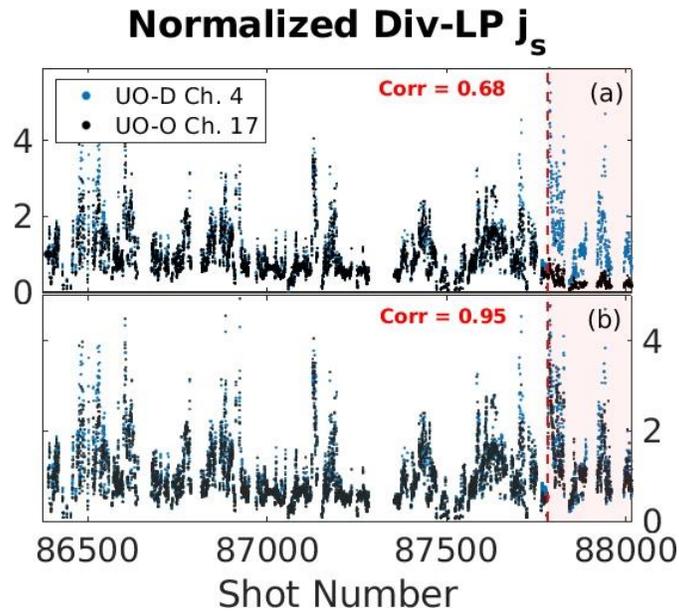

Figure 2 (a) The concatenated $j_s$ measured by a pair of twin channels (the data are normalized to their first elements). (b) The measurements by channel 17 in the red shaded area are multiplied by a constant (mimicking the correction).

The strategy for the correction process is to split the chosen discharges into several parts and correct the less reliable channels for each part in sequence. For each part, a similar table like table 1 (the concatenated $j_s$ data for evaluation of Corr are trimmed for this part) is used to evaluate the partial reliability. All the valid Ohmic discharges in this part are examined to check the measured $j_s$ profiles, the Eich fit, and the performance of the less reliable channels. The correction point for a less reliable channel is determined similarly like channel 17 in figure 2, but the reference channel is not limited to be its twin channel (can also be its neighboring channel if it is reliable). When a correction point is found, its calibration coefficient is corrected following the procedure mentioned above. Figure 3 shows an example of correcting channel 18 in the first part, where all discharges are Ohmic. Different from figure 1, the measurements in figure 3(a) have been corrected. From #86736 to #86868, the $j_s$ measurements by channel 18 are significantly larger than those by its twin channel (channel 5, partial reliable) and neighboring channel (channel 17, overall reliable). The comparisons of the normalized concatenated $j_s$ (trimmed for the first part) with channels 5 and 17 show that the correction point is # 86685. Then the measurement by channel 18 is adjusted using #86736 (see the purple data point in figure 3(c)) and the calibration coefficient is updated and used starting from #86685. The corrected measurements by channel 18 and the new Eich fits are shown in purple to confirm the correction. In the correction process, the significantly biased channel is always corrected firstly to get a relatively trustable Eich fit. To avoid over correction, we don't pursue a perfect Eich fit by correcting all the slightly misaligned reliable channels. It should be noted that only when all the less reliable channels have been corrected, the correction for the next part can be performed.

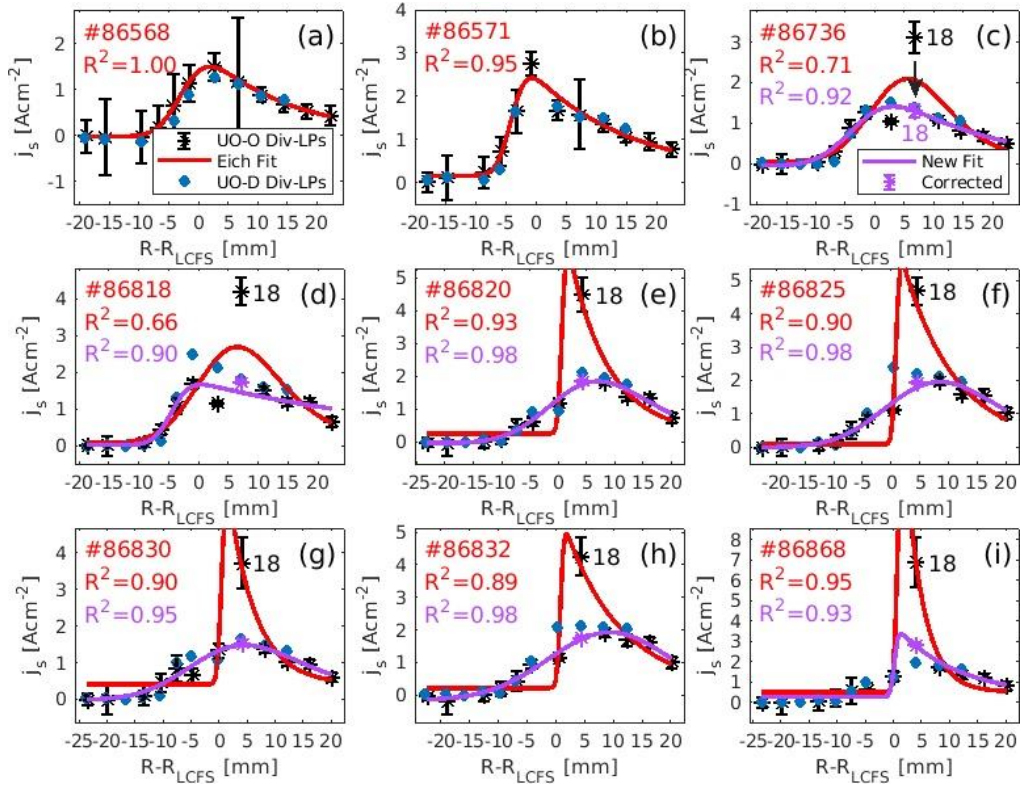

Figure 3 An example to correct the measurements by channel 18 in the first part of the correction process. All the discharges are Ohmic. The measurements by channel 18 increase significantly from #86736 compared with those measured by its more reliable twin and neighboring channels. It is corrected using the Eich fit of the UO-O measurements in #86736. The purple data points and solid lines are the corrected results.

Table 2 lists all the correction points for the UO Div-LPs. Generally, the UO-O Div-LPs have less correction points. Figure 4(a) shows the evolution of the calibration coefficients for the corrected UO-O Div-LP channels. It is seen that there are less correction points for the deuterium plasmas (shaded in grey) and the calibration coefficients change significantly for some of the correction points. There are several possible reasons for the change of the calibration coefficients: i) the measurement stability decreases due to serious erosion of the probe tip; ii) the change of the collecting area due to electromagnetic force for the loose probes; iii) the working state of the power supply changes. No matter what causes the change of the calibration coefficients for the overall less reliable channels, they should be fine to evaluate $\lambda_{js}$ if they can be corrected reasonably. Figure 4(b) compares the coefficient of determination ($R^2$) for the fitting of $j_s$ profiles measured by the UO Div-LPs. The inserted picture shows the cumulative density function of $R^2$. $R^2$ for the UO-O Div-LPs is more concentrated on the $R^2 = 1$ with less correction points (alignment of the measurements with the Eich fit increases $R^2$ but only for corrections in Ohmic discharges). Since the UO-O Div-LPs are more reliable and the key channels of the UO-D Div-LPs are faulty for a long period, the UO-O Div-LPs are used for the evaluation of $\lambda_{js}$ in this paper. Note that it is possible to combine the measurements by the UO-D and UO-O Div-LPs to evaluate $\lambda_{js}$, but the inclusion of less reliable measurements with more correction points might decrease the reliability of $\lambda_{js}$ evaluation and we only assume toroidal symmetry of the UO divertor plasmas for Ohmic discharges. Although the UO-D Div-LPs are not used for correction of the calibration coefficients of the UO-O Div-LPs and the evaluation of $\lambda_{js}$ in this paper, they provide redundant information to assess the channel reliability and find the correction points for the UO-O Div-LPs.

Table 2. The correction points for the UO Div-LPs.

| $N$ | UO-D Correction Points for Ch. $N$ | UO-O Correction Points for Ch. $N+13$ |
|---|---|---|
| 1 | 87782 | |
| 2 | 86685 | |
| 3 | 86685 | |
| 4 | 86378 86568 | 87782 |
| 5 | 86378 86475 86685 86845 86931 87183 | 86572 86685 86844 86884 86931 87127 87250 87782 |
| 6 | 86378 86568 86678 87127 | 86678 86931 |
| 7 | 86568 86818 86931 | 86931 87250 87276 |
| 8 | 86568 86931 87724 | |
| 9 | 86868 87615 87782 | |
| 10 | 86868 | |
| 11 | 86569 86678 86868 | |
| 12 | 86868 | |
| 13 | 86868 | |

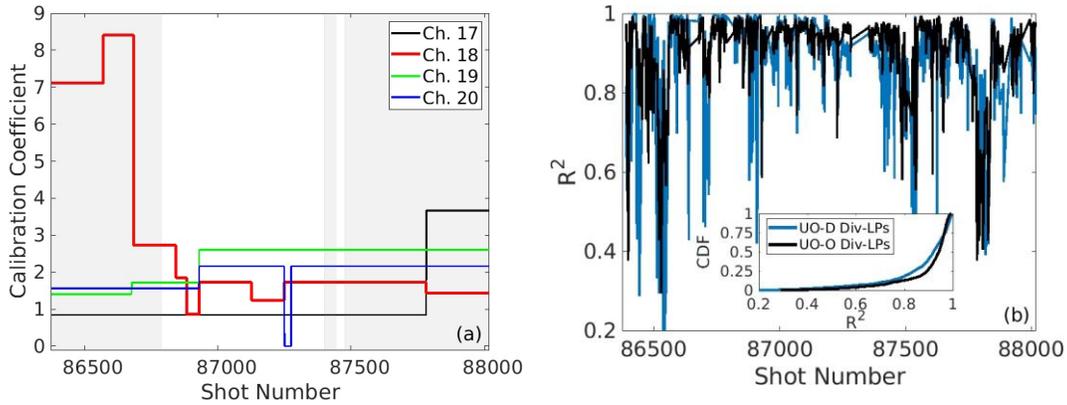

Figure 4 (a) The evolution of the calibration coefficients for the UO-O Div-LPs that have been corrected. The calibration coefficients in the shaded areas are used in this paper. (b) The coefficient of determination ($R^2$) for the fitting of $j_s$ profiles measured by the UO Div-LPs. The inserted picture shows the corresponding cumulative density function (CDF) of $R^2$.

## 4. Scalings of the SOL particle flux width

There are about 900 valid deuterium discharges (see the shaded areas in figure 4(a) for the shot range) for statistical analysis. The filter techniques and the scaling procedures used in this paper are described in reference [7]. About 20 $j_s$ profiles measured by the UO-O Div-LPs are selected and pre-fitted for each of the ~900 discharges. For each of the pre-fitted $j_s$ profile, the selection criteria for a good evaluation of $\lambda_{js}$ is as follows: i) $R^2 \geq 0.88$; ii) the number of main measurement points $n_{main} \geq 4$; iii) the ratio of the minimum value of the SOL measurements to the maximum value of the whole measurements $r_{SOL\text{-}min,peak} < 0.45$; iv) the ratio of the outermost SOL measurement (by channel 14) to the peak-to-peak value of the SOL measurements $r_{SOL,tail} < 0.15$. For each discharge, it is selected if the number of the selected pre-fitted profiles $n_{good} \geq 10$. These selection criteria are important to ensure that the evaluated $\lambda_{js}$ is relatively reliable. Finally, 269 discharges pass the selection. They are further spilt and classified by confinement type (H-mode, L-mode, and Ohmic). For each confinement type, the $j_s$ profiles are averaged every 50 ms, refitted, and selected (manually). The fitting results from 129 H-mode discharges, 103 L-mode discharges, and 32 Ohmic discharges can then be used for constructing the databases.

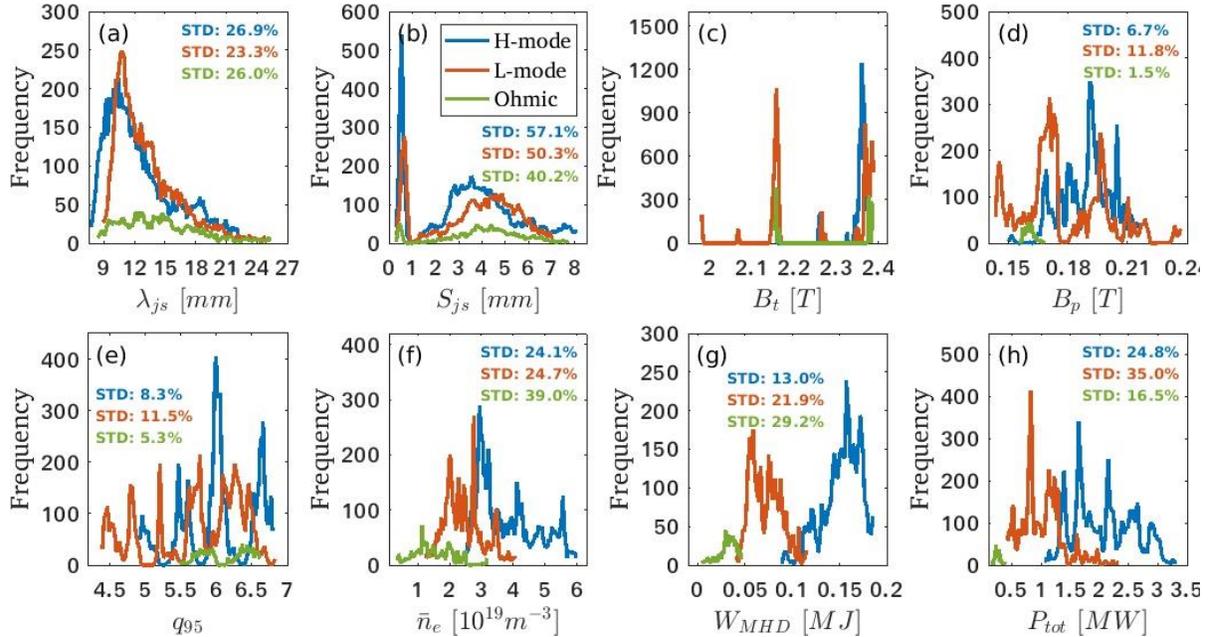

Figure 5 The distributions of the main plasma and engineering parameters for the databases. The standard deviation (STD) is calculated by removing 2% of the data in the head and tail of the distribution.

Three (H-mode, L-mode, and Ohmic) databases are constructed by collecting $\lambda_{js}$, the spreading factor of $\Gamma$ ($S_{js}$), $R^2$, $B_t$, the poloidal magnetic field at LCFS on the OMP ($B_p$), the plasma current ($I_p$), the safety factor at the 95% flux surface ($q_{95}$), $W_{MHD}$, $\bar{n}_e$, $P_{tot}$, the plasma elongation ($\kappa$), the triangularity ($\delta$), etc. The distribution of the main plasma and engineering parameters is shown in figure 5. The distributions of $\lambda_{js}$ peak around $\lambda_{js}$ = 10.5 mm for both H-mode and L-mode databases. The variances for $B_t$, $B_p$, and $q_{95}$ are smaller than those for $\bar{n}_e$, $W_{MHD}$, and $P_{tot}$. The peak values of $\bar{n}_e$, $W_{MHD}$, and $P_{tot}$ distributions increase with respect to the Ohmic, L-mode, and H-mode databases, meeting our expectations. The values of $\kappa$ and $\delta$ for the H-mode and L-mode databases (their values have a small variance less than 5%) are around 1.64, and 0.51, respectively.

We start the power law scaling ($\lambda_{js} = C_0 \prod_{i=1}^{n} x_i^{C_i}$, where $x_i$ is the scaling parameter, $C_i$ is the scaling coefficients, and $i$ is the non-zero natural number) [7] with the parameters that have large variances in figure 5. For the H-mode database, the regression result is listed as #1H in table 3. Since $\lambda_{js}$ has almost no correlation with $q_{95}$ (see table 4), it is removed in #2H. For power law scaling, the co-linearity between the parameters is problematical. One solution is to remove the co-linear parameter that is less correlated with the dependent variable ($\lambda_{js}$). According to table 4, $P_{tot}$ has a relatively large Corr with $\bar{n}_e$ and its correlation with $\lambda_{js}$ is weaker than that for $\bar{n}_e$. So, $P_{tot}$ is removed in #3H ($R^2$ decreases a little bit). For the L-mode database, $P_{tot}$ and $q_{95}$ are stepwise removed in #2L and #3L, respectively, due to their relatively high correlation with $W_{MHD}$. Notice that the absolute values of $W_{MHD}$ and $\bar{n}_e$ are quite close in #3L, these two parameters are combined as $W_{MHD}/\bar{n}_e$ and the result is listed in #4L. Similarly for the H-mode database, #4H is obtained. Observed that #4H and #4L are quite similar, but the parameter range is different (see figure 5), the H-mode and L-mode databases are combined to obtain a unified scaling (#1HL),

$$\lambda_{js,UO} = (2.21 \pm 0.07)(W_{MHD}/\bar{n}_e)^{-0.52\pm0.01}. \qquad (2)$$

Since the Corr between $P_{tot}$ and $W_{MHD}/\bar{n}_e$ is small (0.16), $P_{tot}$ is added back in #2HL ($R^2$ increases a lot),

$$\lambda_{js,UO} = (1.52 \pm 0.04)(W_{MHD}/\bar{n}_e)^{-0.61\pm0.01} P_{tot}^{0.19\pm0.01}. \qquad (3)$$

The detailed scaling results for #1HL and #2HL are shown in figure 6. Compared with the inboard $\lambda_{js}$ scaling (equation (1)), the outboard $\lambda_{js}$ scaling (equation (3)) has stronger scaling dependences on both parameters and a significantly larger scaling amplitude. This in/out asymmetry of $\lambda_{js}$ is possibly explained by Goldston' heuristic drift-based model [11], which is related to the triangularity ($\lambda_{js,UO}/\lambda_{js,UI} = (1 + \delta)/(1 - \delta)$, see equation (9) in reference [7]). Since the scaling exponent of $W_{MHD}/\bar{n}_e$ for equation (2) is nearly the same as that for equation (1), $\lambda_{js,UO}/\lambda_{js,UI,H}$ is approximately equal to 2.21/0.85 = 2.6, which is close to 3 calculated with $\delta$ = 0.5 (see table 3 in reference [7] and the text above). Despite the in/out asymmetry of $\lambda_{js}$, the similar scaling dependence on $W_{MHD}/\bar{n}_e$ confirms that a relatively trustable scaling could be obtained by increasing the amount of data to overcome the relatively large measurement uncertainty by Div-LPs. Still the relatively low $R^2$ might also result from the uncertainties of the EFIT equilibrium and the measurement of $\bar{n}_e$ and/or the absence of important independent scaling parameters.

Table 3. The non-linear regression results of $\lambda_{js}$ using the H-mode and L-mode databases.

| # | $C_0$ | $C_{q_{95}}$ | $C_{\bar{n}_e}$ | $C_{W_{MHD}}$ | $C_{W_{MHD}/\bar{n}_e}$ | $C_{P_{tot}}$ | $R^2$ |
|---|---|---|---|---|---|---|---|
| 1H | 0.62 (0.05) | 0.67 (0.05) | 0.74 (0.02) | -0.31 (0.03) | | 0.34 (0.01) | 0.66 |

| | | | | | | |
|---|---|---|---|---|---|---|
| 2H | 1.55 (0.07) | | 0.67 (0.02) | -0.51 (0.02) | | 0.31 (0.01) | 0.62 |
| 3H | 1.70 (0.08) | | 0.82 (0.02) | -0.47 (0.02) | | | 0.51 |
| 4H | 1.70 (0.08) | | | | -0.62 (0.01) | | 0.48 |
| | | | | | | | |
| 1L | 0.66 (0.03) | 0.63 (0.03) | 0.38 (0.01) | -0.58 (0.02) | | 0.25 (0.01) | 0.61 |
| 2L | 1.35 (0.08) | 0.43 (0.03) | 0.47 (0.01) | -0.41 (0.02) | | | 0.54 |
| 3L | 2.10 (0.10) | | 0.48 (0.01) | -0.52 (0.02) | | | 0.51 |
| 4L | 1.58 (0.08) | | | | -0.60 (0.01) | | 0.54 |
| | | | | | | | |
| 1HL | 2.21 (0.07) | | | | -0.52 (0.01) | | 0.45 |
| 2HL | 1.52 (0.04) | | | | -0.61 (0.01) | 0.19 (0.01) | 0.58 |

Note: The values in the parenthesis are the errors of the scaling coefficients. H, L, and HL in the regression number (#) correspond to the H-mode, L-mode, and combined databases.

According to the results in references [3,4,7], $\lambda_{js}$ is approximately equal to $\lambda_q$ for the Div-LP measurements. Then we can compare the measured $\lambda_{js}$ in the databases with the $\lambda_q$ scalings in previous studies. Figure 7 scales $\lambda_{js}$ against $B_p$ and compares this scaling with the Eich [2] and EAST H-mode [3] $\lambda_q$ scalings. Although the obtained $\lambda_{js}$ scaling (black solid line) has a poor regression quality ($R^2 = 0.18$), its scaling dependence on $B_p$ is almost the same as that for the Eich (red solid line) and EAST (purple solid line) H-mode scalings. However, the scaling amplitude of the $\lambda_{js}$ scaling is ~3 and ~1.6 times larger than that for the Eich and EAST $\lambda_q$ scalings, respectively. The previous explanation for this different scaling amplitude is that the radio-frequency heating scheme in EAST can broaden $\lambda_q/\lambda_{js}$. However, the data from the Ohmic database do not follow the Eich scaling in figure 7, indicating that there exist some unknown broadening mechanisms. Compared with figure 6(a), $\lambda_{js}$ is weakly correlated with $B_p$ in figure 7. This means that the main plasma parameters should be included in the $\lambda_{js}$ scaling to reflect the physics of the SOL heat transports when the variation of $B_p$ is small. Different from the $\lambda_q$ scalings [5] in AUG, the H-mode $\lambda_{js}$ scaling (#4H) is similar to the L-mode $\lambda_{js}$ scaling (#4L) in the scaling amplitude. This might result from that the edge plasma parameters for the H-mode and L-mode plasmas in EAST are similar. Still, the characteristic that the H-mode and L-mode $\lambda_{js}$ scalings can be unified (equations (2-3)) is same as that for the C-Mod ($\lambda_{q,C-Mod} = 0.91\bar{p}^{-0.48}$, where $\bar{p}$ is the volume-averaged core plasma pressure) [6] and AUG ($\lambda_{q,AUG} = 7.57\bar{p}^{-0.52}$) [8] $\lambda_q$ scalings, where the H-mode, I-mode, and L-mode data converge to a single scaling. Furthermore, the scaling exponent of $\bar{p}$ is also nearly the same as that for the scaling exponent of $W_{MHD}$ in the $\lambda_{js}$ scaling ($W_{MHD} \propto \bar{p}$, the plasma volume has a small variation in the databases), which also demonstrates that the obtained inboard and outboard $\lambda_{js}$ scalings by the Div-LPs in EAST are reasonable.

Table 4. The Corr between the scaling parameters for the H-mode and L-mode databases.

| H-mode (L-mode) Corr | $q_{95}$ | $\bar{n}_e$ | $W_{MHD}$ | $P_{tot}$ |
|---|---|---|---|---|
| $\lambda_{js}$ | 0.05 (0.36) | 0.68 (0.51) | -0.29 (-0.46) | 0.56 (0.02) |
| $q_{95}$ | | -0.34 (-0.01) | -0.51 (-0.40) | -0.13 (-0.16) |
| $\bar{n}_e$ | | | -0.02 (0.05) | 0.44 (0.31) |
| $W_{MHD}$ | | | | -0.07 (0.61) |

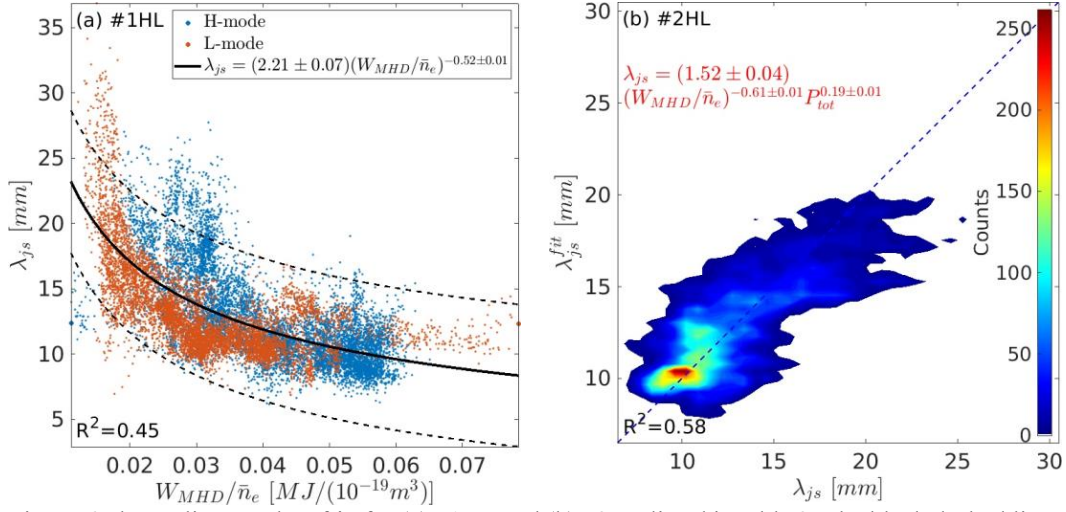

Figure 6 The scaling results of $\lambda_{js}$ for (a) #1HL and (b) #2HL listed in table 3. The black dashed lines represent the scaling uncertainties.

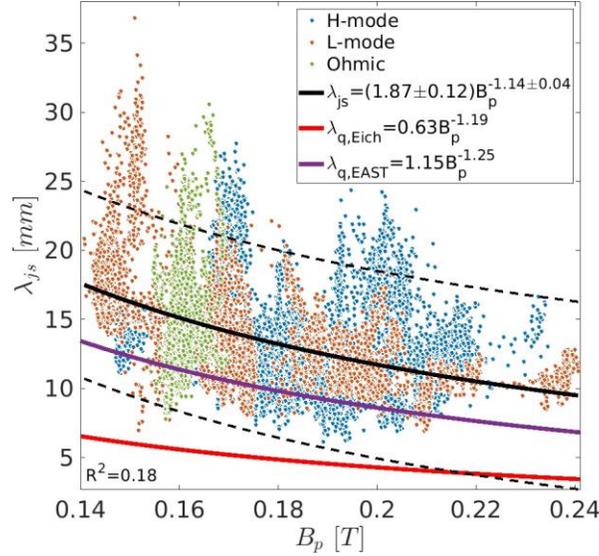

Figure 7 The scaling of $\lambda_{js}$ against $B_p$ with all three databases and the comparisons with the Eich [2] and EAST H-mode [3] $\lambda_q$ scalings. The black dashed lines represent the scaling uncertainties.

## 5. Summaries

In this paper, a systematic method has been developed to correct the less reliable Div-LP measurements by referring the more reliable neighboring and twin channels. The Pearson correlation coefficient is used to evaluate the channel reliability. The correction point for a less reliable channel is determined by comparison of the measurements with its reliable twin and/or neighboring channels. The calibration coefficient is adjusted for the correction point by the Eich fit of the $j_s$ measurements in the Ohmic discharge. According to the correction results, the UO-O Div-LP array is more reliable than the UO-D Div-LP array. It is used for evaluation of the outboard $\lambda_{js}$ in favorable $B_t$ configuration.

About 900 deuterium discharges have been filtered by evaluating the fitting quality of the measured $j_s$ profile, leaving 129 H-mode, 103 L-mode, and 32 Ohmic discharges for constructing the databases. The non-linear regressions of $\lambda_{js}$ have been carried out with respect to the parameters that have large variances for the H-mode and L-mode databases. It is found that the outboard $\lambda_{js}$ scales as, $\lambda_{js,UO} = 1.52(W_{MHD}/\bar{n}_e)^{-0.61}P_{tot}^{0.19}$, for the combined H-mode and L-mode databases. Compared with the inboard $\lambda_{js}$ scaling ($\lambda_{js,UI,H} = 0.85(W_{MHD}/\bar{n}_e)^{-0.50}P_{tot}^{-0.06}$) [7], the outboard scaling has stronger scaling dependences on both parameters but a much larger scaling amplitude. This in/out asymmetry of $\lambda_{js}$ is probably due to triangularity. This scaling confirms that the negative scaling dependence of $\lambda_{js}$ on $W_{MHD}/\bar{n}_e$ is repeatable and trustable despite of that the regression quality is relatively poor. Compared with the typical scaling with respect to $B_p$, the regression quality of the outboard scaling increases, indicating that the solely scaling of $\lambda_q/\lambda_{js}$ with respect to the engineering parameters is not enough to reflect the physics of the SOL heat transports and the key plasma parameters should be included.

## Acknowledgments

This work was supported by the Natural Science Foundation of China (Nos. 12005260, 11922513), and the National Key Research and Development Program (Nos. 2017YFE0301300, 2018YFE0303104). This work was also partially supported by the Institute of Energy, Hefei Comprehensive National Science Center under Grant No. GXXT-2020-004.